\documentclass{article}

\usepackage[english]{babel}
\usepackage{spconf,amsmath,graphicx}
\usepackage{color}
\usepackage{array}
\usepackage{textcomp}
\usepackage{amsmath, amssymb, amsthm}
\usepackage{mathrsfs}
\usepackage{dsfont}
\usepackage{mathabx}
\usepackage{stmaryrd}
\usepackage{csquotes}
\usepackage{enumerate}
\usepackage{booktabs}
\usepackage{float}
\usepackage{graphicx}
\usepackage{xcolor}
\usepackage{hyperref}
\usepackage{enumitem}
\usepackage{txfonts}
\usepackage{comment}

\usepackage[linesnumbered]{algorithm2e}
\SetKwInput{Input}{Input}
\SetKwInput{Output}{Output}

\usepackage{amsthm}

\newtheorem{thm}{Theorem}
\newtheorem{prop}{Proposition}

\newtheorem{lem}{Lemma}

\newcommand{\R}{\mathbb{R}}

\newcommand{\Z}{\mathbb{Z}}
\newcommand{\E}{\mathbb{E}}

\definecolor{brickred}{rgb}{0.8, 0.25, 0.33}

\newcommand{\x}{{\boldsymbol{x}}}
\newcommand{\y}{{\boldsymbol{y}}}
\newcommand{\U}{{\mathbf{U}}}
\newcommand{\Y}{{\mathbf{Y}}}
\newcommand{\X}{{\mathbf{X}}}
\newcommand{\W}{{\mathbf{W}}}
\newcommand{\zero}{{\boldsymbol{0}}}
\newcommand{\XSR}{{\X_\mathrm{SR}}}
\newcommand{\ULR}{{\U_{\mathrm{LR}}}}
\newcommand{\XLR}{{\X_{\mathrm{LR}}}}

\newcommand{\tXLR}{{\tX_\mathrm{LR}}}
\newcommand{\Uref}{{\U_\mathrm{ref}}}
\newcommand{\tX}{\tilde{{\X}}}
\newcommand{\bb}{{\mathbf{b}}}
\newcommand{\A}{{\mathbf{A}}}
\newcommand{\B}{{\mathbf{B}}}
\newcommand{\bC}{{\mathbf{C}}}
\newcommand{\Sub}{{\mathbf{S}}}
\newcommand{\G}{{\boldsymbol{\Gamma}}}
\newcommand{\La}{\boldsymbol{\Lambda}}
\newcommand{\la}{\boldsymbol{\lambda}}
\newcommand{\ka}{{\boldsymbol{\kappa}}}
\newcommand{\be}{\boldsymbol{\beta}}
\newcommand{\bc}{\mathbf{c}}
\newcommand{\bt}{\mathbf{t}}
\newcommand{\g}{\boldsymbol{\gamma}}
\newcommand{\OMN}{{\Omega_{M,N}}}
\newcommand{\OMNr}{{\Omega_{M/r,N/r}}}
\newcommand{\OMNklr}{{\Omega^{k,\ell,r}_{M,N}}}
\DeclareMathOperator{\Tr}{Tr}
\DeclareMathOperator{\ADSN}{ADSN}
\DeclareMathOperator{\per}{\textbf{per}}

\title{Stochastic super-resolution for Gaussian textures}
\name{\'Emile Pierret$^a$, Bruno Galerne$^{a,b}$}
\address{$^a$ Institut Denis Poisson -- Université d'Orléans, Université de Tours, CNRS\\
$^b$ Institut universitaire de France (IUF)}

\begin{document}
\ninept

\maketitle
\begin{abstract}
Super-resolution (SR) is an ill-posed inverse problem which consists in proposing high-resolution images consistent with a given low-resolution one.
While most SR algorithms are deterministic, 
stochastic SR deals with designing a stochastic sampler generating any realistic SR solution.
The goal of this paper is to show that stochastic SR is a well-posed and solvable problem when restricting to Gaussian stationary textures. 
Using Gaussian conditional sampling and exploiting the stationarity assumption, we propose an efficient algorithm based on fast Fourier transform.
We also demonstrate the practical relevance of the approach for SR with a reference image. 
Although limited to stationary microtextures, our approach compares favorably in terms of speed and visual quality to some state of the art methods designed for a larger class of images.
\end{abstract}
\section{Introduction}
\label{sec:intro}

The super resolution (SR) problem consists in generating a high-resolution (HR) image corresponding to a given low-resolution (LR) input image. This is a very ill-posed inverse problem that necessitates an image prior model or additional information to create images having sharp edges and rich texture details. This is a very important problem for the entertainment industry due to the increase in resolution of display screens as well as other imaging industries where resolution is critical, notably satellite imaging and microscopy.

Different frameworks are considered in the literature: the Single Image SR (SISR) consists in using only an LR input image and a generic database of HR images, while SR with a reference image considers an LR input image accompanied with an HR image that presents similarity with the unknown HR input.
Most contributions in SISR are based on conditional generative neural networks \cite{Isola_et_al_2017, Bruna_et_al_2016, Ledig_et_al_2017, Wang_et_al_2018}, adversarial \cite{Goodfellow_et_al_2014}  or not, and trained using the “perceptual loss” \cite{Johnson_et_al_2016}, that is, the distance between pretrained VGG features \cite{Simonyan_Zisserman_2014}.
These references tackle zoom factor x4 or even x8. 
Contrary to a classical x2 problem that can be seen as an image sharpener problem, 
for such high zoom factors new image content must be generated in accordance with the LR input, and the space of possible images becomes very large. 
All references then insist on the importance of generating local textures and avoiding the “regression to the mean problem”: 
when favoring PSNR the optimal result consists in a blurry image close to the mean of plausible sharp images.
While SISR can be seen as local conditional texture synthesis, a close inspection of the state-of-the-art techniques shows however that texture modeling is absent from GANs. 
Indeed, the perception loss does not favor the statistics of the textures one wishes to reconstruct. 
In addition, the proposed models are generally deterministic, which is not desirable for texture generation since it does not allow checking the statistical consistency of the proposed solution. 
More recent models propose generative networks with stochastic response for SISR: SRFlow \cite{Lugmayr_et_al_2020} use invertible generative flow \cite{Kingma_Dhariwal_2018} and \cite{Saharia_et_al_2021} use Denoising Diffusion Probabilistic Models (DDPM) \cite{Ho_et_al_2020}, also called score-based models \cite{Song_Ermon_2019}.
They propose practical solutions for stochastic SR, that is providing a stochastic sampler that outputs any realistic solution to the SISR problem.
In addition, let us note that several recent contributions solely focus on texture SR, using patch statistics from a reference image \cite{Hertrich_et_al_Wasserstein_patch_prior_superresolution_IEEETCI2022, Nguyen_et_al_2022} or a prior as spectral content \cite{chatillon_statistically_2022}.

In this paper, we solve the problem of stochastic SR when restricting to stationary Gaussian textures \cite{Galerne_Gousseau_Morel_random_phase_textures_2011}. Following a similar approach as for Gaussian texture inpainting~\cite{Galerne_Leclaire_Moisan_microtexture_inpainting_icassp2016, Galerne_Leclaire_gaussian_inpainting_siims2017},
assuming that the LR input is stationary and Gaussian with a known covariance in HR space, we propose an exact SR sampler relying on conditional Gaussian simulation.
We exploit the stationarity of both the texture model and the zoom-out operator to obtain an efficient algorithm based on Fast Fourier Transform (FFT).
We further demonstrate experimentally the interest of the approach in the context of SR with a reference image.

The plan of the paper is as follows. 
We present our framework of SR and remind results about conditional Gaussian simulation. Then, we focus on Gaussian stationary textures and detail how to implement efficiently the stochastic simulation for these models. 
Finally, we extend the approach for SR with a reference image and compare our results with the state of the art before concluding.

\begin{figure*}[htb]
    \centering
    {
    \begin{minipage}[b]{0.196\textwidth}
    \centerline{ LR image}
        \includegraphics[width=\textwidth]{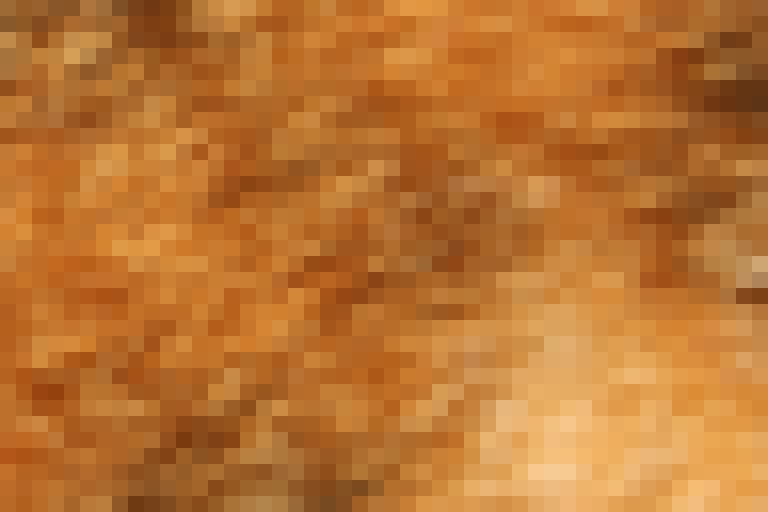}
    \end{minipage}
    \begin{minipage}[b]{0.196\textwidth}
    \centerline{ HR image}
        \includegraphics[width=\textwidth]{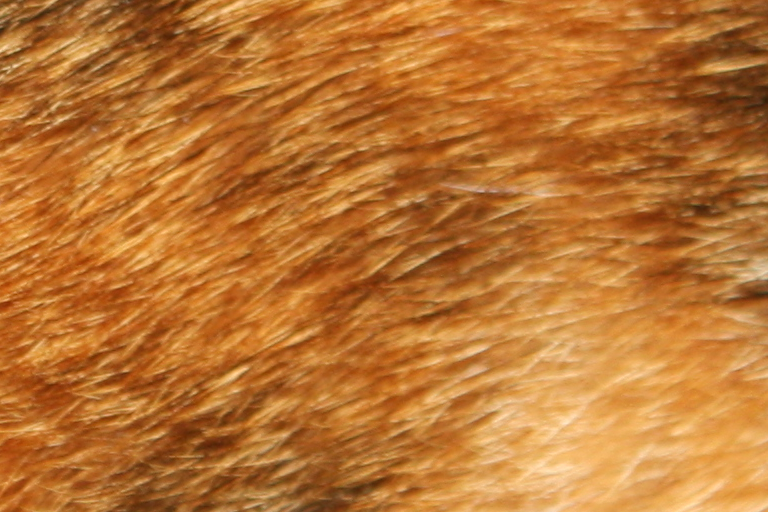}
    \end{minipage}
        \begin{minipage}[b]{0.196\textwidth}
        \centerline{ SR Gaussian sample}
        \includegraphics[width=\textwidth]{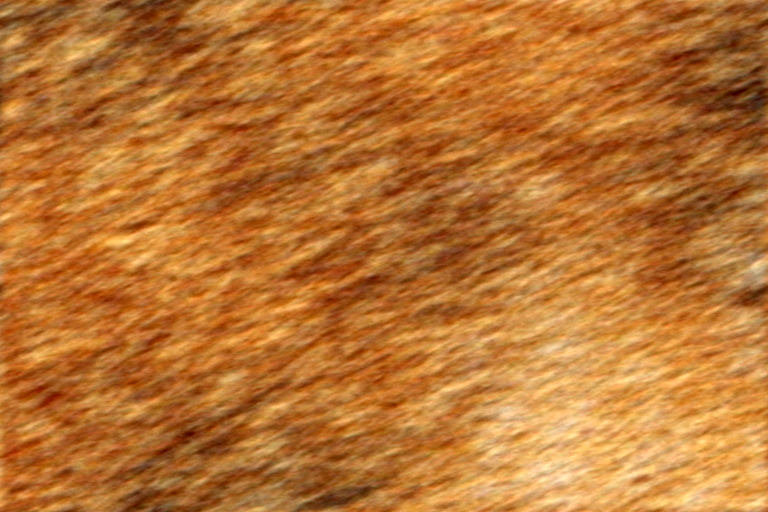}
    \end{minipage}
        \begin{minipage}[b]{0.196\textwidth}
        \centerline{ Kriging component}
        \includegraphics[width=\textwidth]{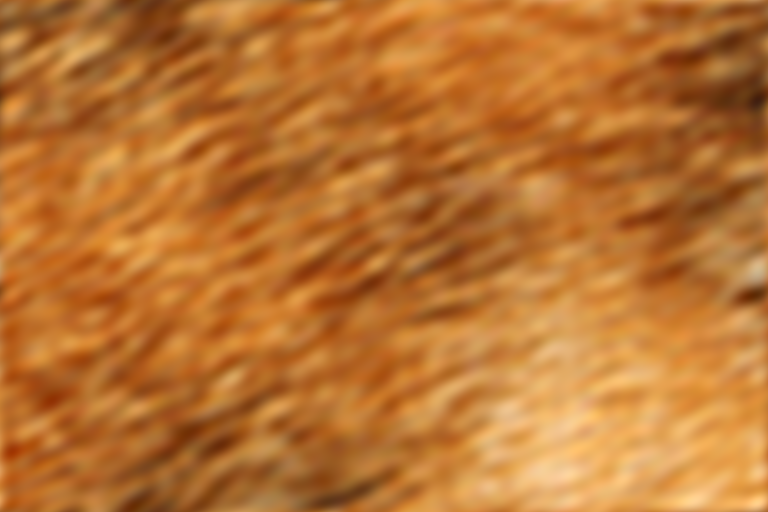}
    \end{minipage}
        \begin{minipage}[b]{0.196\textwidth}
        \centerline{ Innovation component}
        \includegraphics[width=\textwidth]{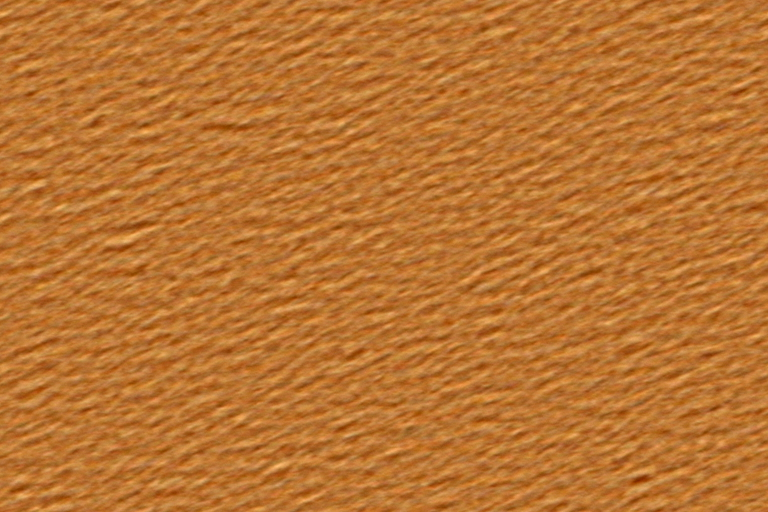}
    \end{minipage}  \\
    \vspace{0.003\textwidth}
    \begin{minipage}[b]{0.196\textwidth}
        \includegraphics[width=\textwidth]{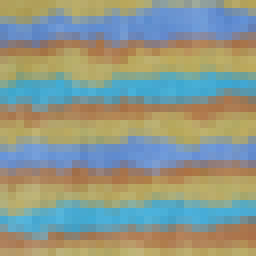}
    \end{minipage}
    \begin{minipage}[b]{0.196\textwidth}
        \includegraphics[width=\textwidth]{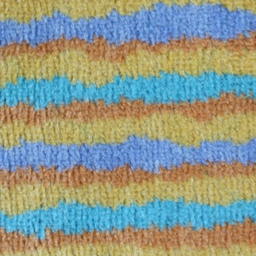}
    \end{minipage}
        \begin{minipage}[b]{0.196\textwidth}
        \includegraphics[width=\textwidth]{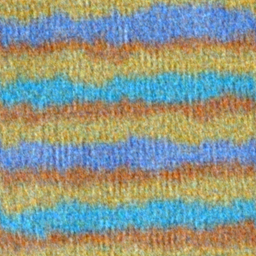}
    \end{minipage}
        \begin{minipage}[b]{0.196\textwidth}
        \includegraphics[width=\textwidth]{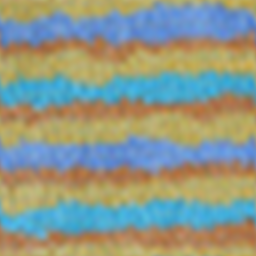}
    \end{minipage}
        \begin{minipage}[b]{0.196\textwidth}
        \includegraphics[width=\textwidth]{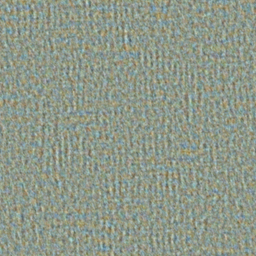}
    \end{minipage}%
    }%
    \caption{\label{fig:sr_gaussian}
    Different components of the Gaussian SR model. 
    Top row: SR with a factor $\times 16$, the images HR size is $512 \times 768$;
    Bottom row: SR with a factor $\times 8$, the images HR size is $256 \times 256$.
    The last columns present respectively a Gaussian SR sample and its associated kriging and innovation components. 
    Observe the complementarity of the the kriging component which deblurs the LR input and the innovation component which adds independent high-frequency details.
    }
\end{figure*}

\section{Gaussian conditional simulation for super-resolution}

\subsection{Framework}

\textbf{Notations} 
Let $M,N > 2$ be the size of the images. We extend all the images on $\Z^2$ by  periodization.
We write $[M] = \{0,\ldots,M-1\}$ and $\OMN = [M]\times[N]$. 
For $\X,\Y \in \R^{\OMN}$, $\X \star \Y$
designates the discrete and periodic convolution defined by \mbox{$
    (\X \star \Y)(\x) =  \sum_{\y \in \OMN} \X\big (\x-\y\big)\Y(\y)$}, $\x \in \OMN$. 
For $\la \in \R^{ \OMN}$, we denote the symmetric $\widecheck{\la}$ such that for $\x \in  \OMN, \widecheck{\la}(\x) = \la(-\x)$. 
For $\X \in \R^{\OMN}$, we express by $\mathscr{F}_2(\X)$ or $\widehat{\X}$ 
its discrete Fourier transform defined by
$\widehat{\X}(\x) =   \sum_{\y \in \OMN} \X(\y) \exp({-\frac{2i\pi x_1 y_1}{M}})\exp(-\frac{2i\pi x_2 y_2}{N})$, $\x \in \OMN$. Let us recall that $\mathscr{F}_2$ is invertible with $\mathscr{F}^{-1}_2 = \frac{1}{MN}\overline{\mathscr{F}_2}$ and that $\mathscr{F}_2(\X \star \Y) = \mathscr{F}_2(\X)\odot \mathscr{F}_2(\Y)$ where  $\odot$ denotes the componentwise product.

\noindent \textbf{Zoom-out operator}
We denote by $r$ the factor of zoom-out $ ( = 4,8, \ldots )$. From now on we assume that $M$ and $N$ are multiples of $r$. The loss of pixels in an image can be caused by multiple operators. We will assume in our experiments that the degradation is caused by the imresize function of Matlab which is used in many models~\cite{Ledig_et_al_2017, Wang_et_al_2018, Lugmayr_et_al_2020}.
Concretely, this operator reduces the size of an image by weighting the pixels. 
Supposing that the image is periodic, we can split it off in a separable convolution which weights all the pixels followed by a subsampling with stride $r$ \cite{bahat2020explorable}. 
We denote $\A = \Sub\bC$ this linear zoom-out operator, with $\bC$ the convolution and $\Sub$ the subsampling operator with stride $r$ such that for $\U \in \R^{\OMN}, \Sub\U \in \R^{\OMNr}$ and for $\x \in \OMNr$, $(\Sub\U)(\x) = \U(r\x)$.

\subsection{Gaussian conditional simulation}

Let $\X \in \R^{\OMN}$ be a stationary Gaussian process with distribution $\mathscr{N}(\zero,\G)$ and $\XLR = \A\X \in \R^{\OMNr}$ be its LR version.
$\A$ reduces the size of the images and thus is not invertible. 
We would like to sample $\X | \XLR$. In our further assumptions, $\X$ is supposed to be Gaussian, consequently we will use the classical theorem recalled in \cite{Galerne_Leclaire_gaussian_inpainting_siims2017}:
\begin{thm}[Gaussian simulation and Gaussian kriging]
\label{thm:kriging}
Let $\X$ be a \textbf{Gaussian} vector, and $\A$ be a linear operator,
$\E(\X|\A\X)$ and $\X - \E(\X|\A\X)$ are independent. Consequently, if $\tX$ is independent of $\X$ with the same distribution, then $\E(\X|\A\X) + [\tX - \E(\tX|\A\tX) ]$ has the same distribution as $\X$, knowing $\A\X$.
Furthermore, if $\X$ is zero-mean, there exists $\La \in \R^{{\OMNr}\times{\OMN}}$ such that $\E(\X|\A\X) = \La^T\A\X$ and $\E(\X|\A\X) = \La^T\A\X$ if and only if $\La$ verifies the matrix equation:
  \begin{equation}
    \label{eq:kriging_matrix}
    \A\G \A^T\La = \A\G.
  \end{equation}
\end{thm}

\noindent This theorem is a consequence of the fact that the space of square-integrable random variables is a Hilbert space. The assumption that $\X$ follows a Gaussian law is crucial. As for now, the resolution of our problem requires the computation of  $ \E(\X|\XLR) =\E(\X|\A\X) = \La^T\A\X = \La^T\XLR$, where $\La$ is called the kriging matrix.  $\La^T\XLR$  is called the kriging component and $\tX - \La^T \A\tX$ the innovation component. Therefore, stochastic SR for Gaussian textures corresponds to simulate $\La^T\XLR + (\tX-\La^T \A \tX)$. 
Figure \ref{fig:sr_gaussian} presents the realization of $\X|\XLR$, titled SR Gaussian  sample. This is the sum of the kriging component which is an attachment to data and the innovation component that adds stochastic independent high-frequency variations.
For the given LR image $\XLR$ and the HR covariance $\G$, it is just necessary to compute $\La$ to sample HR versions of $\XLR$

\section{An efficient implementation in the stationary case}

\subsection{ADSN model and super-resolution}

Our objective is to study the SR of Gaussian stationary textures. 
We will focus on textures following the asymptotic discrete spot noise (ADSN) model~\cite{Galerne_Gousseau_Morel_random_phase_textures_2011}. 
Given a grayscale image $\U\in\R^{\OMN}$ with mean grayscale $m\in\R$,
one defines $\ADSN(\U)$ as the distribution 
of $\X = \bt \star \W$
where $\bt = \frac{1}{\sqrt{MN}}(\U-m)$ is called the texton associated with $\U$.
$\ADSN(\U)$ is a Gaussian distribution with mean zero and covariance matrix $\G$ that represents the convolution by the kernel $\g = \bt \star \widecheck{\bt}$.
$\X\sim \ADSN(\U)$ is stationary since for all $\x \in \OMN$ and $\y \in \Z^2$, $\X(\x - \y) \overset{\mathscr{L}}{=} \X(\x)$.

For a given input image $\U \in \R^{\OMN}$ and its LR version $\ULR = \A\U$, we would like to sample 
$\X\sim\ADSN(\U)$ conditioned on $\XLR=\ULR$, that is, by Theorem~\ref{thm:kriging},
\begin{equation*}
\XSR = \La^T \ULR + (\tX - \La^T \A\tX)
\end{equation*}
where $\tX\sim \ADSN(\U)$, and $\La$ is the kriging matrix associated with $\ADSN(\U) = \mathscr{N}(\zero,\G)$. We will explain how to implement it efficiently in the following.

\subsection{Structure of the problem induced by stationarity}

For $k,\ell \in [r]$, let 
$\OMNklr = \{(k + ir,\ell + jr), ~ i,j \in [M/r]\times[N/r] \}\subset \OMN$ be the subgrid of $\OMN$ having stride $r$ and starting at $(k,\ell)$. 
Note that each subgrid $\OMNklr$ has the same number of pixels as the LR image domain $ \OMNr$.

\begin{prop}[Structure of the kriging matrix] \label{prop:structure_kriging_matrix}
There exists $\La \in \R^{{\OMNr}\times{\OMN}}$ solution of Equation~\ref{eq:kriging_matrix} such that $\Y \in \R^{\OMNr}\mapsto \La^T \Y \in \R^{\OMN}$ corresponds to a convolution on each of the shifted subgrids $\OMNklr$, $k,\ell \in [r]$. 
More precisely, 
$\La$ is fully determined by its $r^2$ first columns $\la(k,\ell) = \La_{\OMNr\times (k,\ell)}$, $k,\ell \in [r]$ and 
$$
\left(\La^T \Y\right)(\OMNklr) = \widecheck{\la}(k,\ell) \star \Y.
$$
\end{prop}

\noindent The proof of Proposition~\ref{prop:structure_kriging_matrix} relies on the fact that both $\X$ and $\XLR = \A\X$ are stationary on their respective domains.
Consequently, it is sufficient to solve the $r^2$ equations
  \begin{equation}
    \label{suffeqkrig}
    \A\G\A^T\la(k,\ell) = \A\G_{\OMN\times (k,\ell)}, \quad k,\ell \in [r]
    %\tag*{$(2)_{k,\ell}$}
  \end{equation}
to fully determine the kriging component of $\X|\XLR$.
Remark that the matrix $\La$ is determined by $MN$ values, the size of the HR image.

\subsection{Resolution of the systems}

To determine the kriging component, it is sufficient to solve the $r^2$ systems of Equation \eqref{suffeqkrig}.
We will use the Lemma \ref{lem:convolution_subsampling}.

\begin{lem}[Convolution and subsampling]
\label{lem:convolution_subsampling}
If $\B$ is a 2D convolution on $\OMN$ by the kernel $\be \in \OMN$, $\Sub\B\Sub^T $ is a convolution on $\OMNr$ by the kernel $\Sub\be \in \OMNr$.
\end{lem}

\noindent By Lemma \ref{lem:convolution_subsampling}, $\A\G\A^T$ is a convolution matrix with kernel $\ka = \Sub(\bc \star \g \star \widecheck{\bc})$ where $\bc$ is the kernel of $\bC$. The Equations \eqref{suffeqkrig} become
  \begin{equation}
    \ka \star \la(k,\ell) = \A\G_{\OMN\times (k,\ell)}, \quad k,\ell \in [r]
  \end{equation}
and we can easily pseudo-invert the convolution by component-wise division in the Fourier domain. 

The full FFT-based procedure is given by Algorithm~\ref{algo:gaussian_sr}.
Note that to avoid artefacts due to non-periodicity of the images, as classically done, we use the periodic plus smooth decomposition~\cite{moisan_periodic_2011} to make the spectral computations. To work with non-zero mean images, the mean grayscale of the LR image should be subtracted at the beginning of the algorithm and added at the end. For RGB color images, the operations should be led on each channel. Note that the Gaussian noise $\W$ to generate the ADSN textures should be the same for each channel~\cite{Galerne_Gousseau_Morel_random_phase_textures_2011}, as done for the results prensented in Figure~\ref{fig:sr_gaussian}.

\setlength{\textfloatsep}{0.1cm}

\begin{algorithm}[t!]

{ \Input{An image $\ULR \in \R^{\OMNr}$, $r$ the zoom factor, $\bt$ the convolution kernel of the ADSN model, $\bc$ the kernel of the convolution of the zoom-out operator $\A = \Sub\bC$}
{\textit{\textbf{Step 1:} Computation of kriging matrix $\La$}}\:

{Store $\per(\bt)$ the periodic component of $\bt$}\;

{Store the convolution kernels $\g = \per(\bt)\star \textbf{p$\widecheck{\textbf{e}}$r}(\bt)$, $\bc \star \g$ and $\ka = \bc \star \g \star \widecheck{\bc}$ (computed in Fourier)}\;

\For{$(k,\ell)\in[r]^2$}{
{$\widehat{\bb} = \mathscr{F}_2\Big(\Sub((\bc \star \g)(\cdot-k,\cdot-\ell))\Big)$ \label{rollcg}}\;

{$\widehat{\la}(k,\ell)\left[\widehat{\ka}\neq 0\right] \leftarrow$  {\Large$\frac{\widehat{\bb}[\widehat{\ka} \neq 0]}{\widehat{\ka}[\widehat{\ka} \neq 0]}$}}\;}

{\textit{\textbf{Step 2:} Sampling of one SR version of $\ULR$}}\;
{Generate $\W \in \R^{\OMN}$ following a Gaussian standard law}\;
{$\tX \leftarrow \bt \star \W$}\;
{$\tXLR \leftarrow \A \tX$}\;

\For{each shifted subgrid by $(k,\ell)\in[r]^2$}{
{$\XSR(\OMNklr)  \leftarrow \mathscr{F}_2^{-1}\left((\widehat{\U}_\mathrm{LR} - \tXLR) \odot \overline{\widehat{\la}(k,\ell)}\right) + \tX(\OMNklr)$}\;}

\Output{$\XSR$}
\BlankLine
\caption{\label{algo:gaussian_sr} Pseudo-code of the Gaussian simulation for SR. To generate several samples, only the step 2 should be re-run.}}
\end{algorithm}

\section{Super-Resolution with a reference image}

\newlength{\srrefwidth}
\setlength{\srrefwidth}{0.165\textwidth}

\begin{figure*}[ht!]
\centering
\setlength{\tabcolsep}{0.5pt}
\setlength{\extrarowheight}{0.1pt}
\begin{tabular}{@{}*{6}{c}@{}}
LR image
&
Reference image
&
HR image
&
Kriging component
&
Bicubic
&
SRFlow ($\tau = 0$)
\\
\includegraphics[width=\srrefwidth]{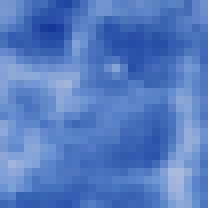}
& 
\includegraphics[width=\srrefwidth]{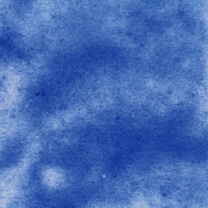}
& 
\includegraphics[width=\srrefwidth]{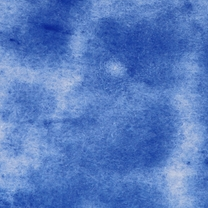}
&
\includegraphics[width=\srrefwidth]{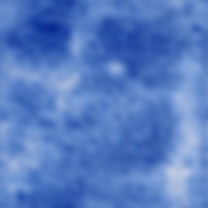} 
&
\includegraphics[width=\srrefwidth]{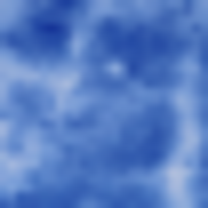}
&
\includegraphics[width=\srrefwidth]{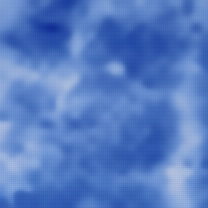}
\\

\multicolumn{3}{c}{Evaluation metrics} 
&
Gaussian SR (ours)
&
WPP
&
SRFlow ($\tau = 0?9$)
\\
 \multicolumn{3}{l}{
 \raisebox{1.1\height}{\begin{minipage}[b]{0.495\textwidth}
 {\footnotesize
    \begin{tabular}{l*{5}{l}}
    \toprule
    & PSNR (dB) $\uparrow$  & SSIM $\uparrow$  & LPIPS $\downarrow$ & TIME (s)  \\
    \midrule
    Kriging component \text{  } & 28.42 & \underline{0.56} & 0.69 &  \\ \hline
Bicubic & $\mathbf{30.21}$ & $\mathbf{0.65}$ & 0.54 &  \\ \hline
$\text{SRFlow}{(\tau = 0)}$  & $\underline{28.55}$ & 0.54 & 0.63 & \underline{0.47} (GPU) \\ \hline
Gaussian SR (ours)  & \text{ } 26.25 $\pm$ 0.05 \text{ } & 0.42 $\pm$ 0.00 & $\mathbf{0.12 \pm 0.01}$ & $\mathbf{0.01}$ (CPU) \\ \hline
WPP & 24.70 & 0.39  & 0.22 & 64.0 (GPU) \\ \hline
$\text{SRFlow}{(\tau = 0.9)}$   & \text{ } 27.33 $\pm$ 0.34 \text{  } & 0.48 $\pm$ 0.02 \text{ } & \underline{0.20 $\pm$ 0.03} \text{ } & \underline{0.47} (GPU)\\
\bottomrule
\end{tabular}}
\end{minipage}}}

&
\includegraphics[width=\srrefwidth]{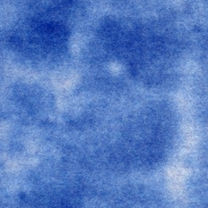}

& 
\includegraphics[width=\srrefwidth]{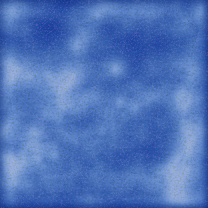}

& 
\includegraphics[width=\srrefwidth]{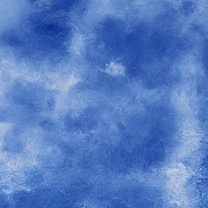}
\\
LR image
&
Reference image
&
HR image
&
Kriging component
&
Bicubic
&
SRFlow ($\tau = 0$)
\\

\includegraphics[width=\srrefwidth]{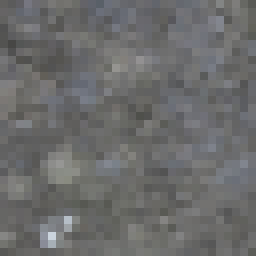}

& 
\includegraphics[width=\srrefwidth]{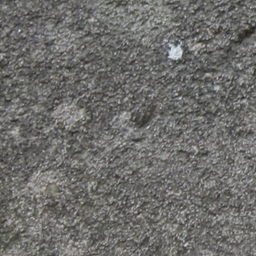}

& 
\includegraphics[width=\srrefwidth]{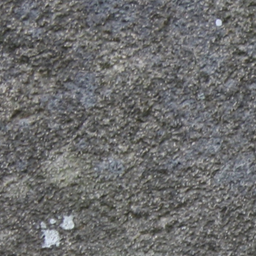}
& 
\includegraphics[width=\srrefwidth]{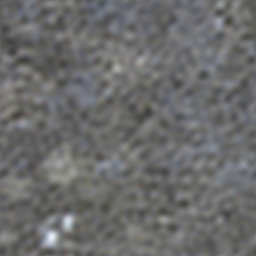}

& 
\includegraphics[width=\srrefwidth]{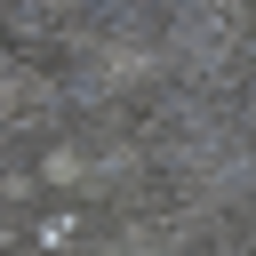}
&
\includegraphics[width=\srrefwidth]{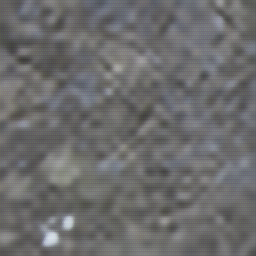}
 \\
\multicolumn{3}{c}{Evaluation metrics} 
&
Gaussian SR (ours)
&
WPP
&
SRFlow ($\tau = 0?9$)
\\
 
\multicolumn{3}{c}{
\raisebox{1.1\height}{\begin{minipage}[b]{0.495\textwidth}
{\footnotesize
    \begin{tabular}{l*{5}{l}}
    \toprule
     & PSNR (dB) $\uparrow$  & SSIM $\uparrow$  & LPIPS $\downarrow$ & TIME (s) \\
     \midrule
     Kriging component \text{  } & 21.78 & \underline{0.24} & 0.87 & \\ \hline
     Bicubic  & $\mathbf{23.52}$ & $\mathbf{0.45}$ & 0.70 & \\ \hline
$\text{SRFlow}{(\tau = 0)}$   & \underline{21.84} & \underline{0.24} & 0.87 & \underline{0.55} (GPU) \\ \hline
Gaussian SR  (ours) & 18.99 $\pm$ 0.05 \text{  } & 0.14 $\pm$ 0.01\text{  }  & $\mathbf{0.25 \pm 0.01}$ \text{  } & $\mathbf{0.02}$ (CPU) \\
\hline
 WPP & 21.12 & 0.21 & 0.42 &  77.0 (GPU) \\ \hline
$\text{SRFlow}{(\tau = 0.9)}$   & 18.99 $\pm$ 0.38 & 0.14 $\pm$ 0.01 & \underline{0.39 $\pm$ 0.04} & \underline{0.55} (GPU) \\ 
\bottomrule
    \end{tabular}
    }
\end{minipage}}}
&
\includegraphics[width=\srrefwidth]{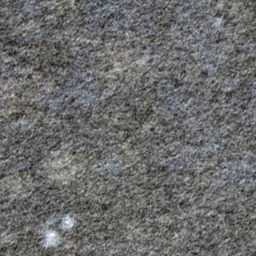}
&
\includegraphics[width=\srrefwidth]{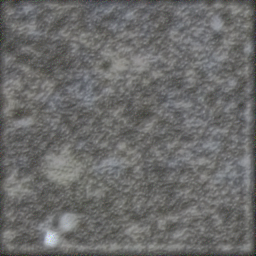}
&
\includegraphics[width=\srrefwidth]{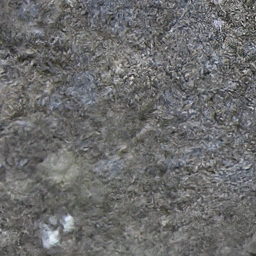}

\end{tabular}
\caption{ Gaussian SR with a reference image for a factor $\times 8$ and comparison with other methods.
Top: HR size $208 \times 208$. Bottom:
HR size $256 \times 256$. For the stochastic methods SR Gaussian and SRFlow, the table has been realized on 200 samples. Note that our method outperforms in terms  of the perceptual LPIPS metric and execution time, while PSNR and SSIM are optimal for blurry images, illustrating that these metrics are not relevant for texture SR. Images from~\cite{ggm_rpn_ipol2011}.\label{fig:Gaussian_sr_with_ref}}
\end{figure*}

In the previous theoretical sections we used the ground truth HR version $\U$ to estimate the distribution of the Gaussian texture, making the process impractical.
In this section we demonstrate that the approach extends to the context of SR with a reference image, the reference image being used in place of the ground truth for modeling the ADSN distribution.
Given an LR texture $\ULR = \A\U \in \R^{\OMNr}$ and a reference HR texture $\Uref \in \R^{\OMN}$, 
we simply replace the unkown ADSN kernel $\bt = \frac{1}{\sqrt{MN}}(\U-m)$ by $\bt_\mathrm{ref} = \frac{1}{\sqrt{MN}}(\Uref-m_\mathrm{ref})$ in Algorithm~\ref{algo:gaussian_sr}, assuming that the two microtextures are similar.
We sample $\XSR = \La^T \ULR+ (\tX -  \La^T \A\tX)$ where $\tX$ has the distribution $\ADSN(\Uref)$ and $\La$ is the kriging matrix associated with $\ADSN(\Uref)$.
Two SR experiments are given in Figure~\ref{fig:Gaussian_sr_with_ref} where we evaluate and compare our results with two competitive methods: Wasserstein patch prior (WPP)~\cite{Hertrich_et_al_Wasserstein_patch_prior_superresolution_IEEETCI2022} and SRFlow~\cite{Lugmayr_et_al_2020}.
WPP\footnote{\ninept\url{https://github.com/johertrich/Wasserstein_Patch_Prior}} is deterministic and proposes to use optimal transport in patch space in a variational formulation for SR. 
The routine needs a reference image similar to ours. 
SRFlow\footnote{\ninept Code and weights from \url{https://github.com/andreas128/SRFlow}} is a normalizing flow for sotchastic SR trained on natural images.
The network transforms a standard Gaussian latent variable into a SR sample given the LR image. 
It depends on a temperature hyperparameter $\tau \in [0,1]$ that modulates the variance of the latent variable. 
Both methods require GPU and experiments were conducted using a NVIDIA V100 GPU.
Visually, there are some issues in the borders of the image generated by WPP and the output texture tends to be blurry. 
The results of SRFlow with high temperature are too sharp for the Gaussian texture input, probably a bias induced by the training set of natural images that are often piecewise regular.
The images obtained with SR Gaussian (ours) have a more granular appearence, due to the innovation component. However, the texture of details that breaks the stationary assumptions are not retrieved as in the white spot in the second image.
Let us stress here that both WPP and SRFlow are designed for generic natural images while our method only works on Gaussian textures.

To evaluate the different methods we report in Figure~\ref{fig:Gaussian_sr_with_ref} the metrics Peak Signal to Noise Ratio (PSNR), Structural SIMilarity (SSIM) and Learned Perceptual Image Patch Similarity (LPIPS). 
PSNR is the logarithm of the mean-squared error (MSE) between two images. 
SSIM is a metric which quantifies the similarity of two images studying their luminance, contrast and structure~\cite{wang_image_2004}. 
LPIPS is the norm between weighted features of a pre-trained classification network and quantifies the perceptual similarity between two images~\cite{zhang_unreasonable_2018}. 
As illustrated by the results of Figure~\ref{fig:Gaussian_sr_with_ref},
PSNR and SSIM are not adapted to evaluate the quality of the texture samples in our context.
Indeed the best solutions for this metric are always the most blurry solutions, namely the bicubic interpolation, SRFlow with temperature $\tau=0$, and the kriging component.
The LPIPS metric is more relevant for SR of textures. 
For this metric our Gaussian SR results are the best for both examples.
Note also that our algorithm runs very quickly using only a CPU.

To advocate for the irrelevance of the PSNR for stochastic texture SR, in our Gaussian framework it can be shown that the MSE is lower for the blurry kriging component than for the perfect SR samples.

\begin{prop}[Kriging component and MSE]
\label{prop:MSE}
Let $\U \in \R^{\OMN}$ be a Gaussian texture, $\ULR = \A\U$ its LR version, $\Uref$ a reference image,  $\G$ such that $\ADSN(\Uref) = \mathscr{N}(\zero,\G)$ and $\La$ associated to $\G$.
Let $\XSR$ be a random image following the distribution of the SR samples obtained with Algorithm \ref{algo:gaussian_sr}, then
\begin{equation*}
\begin{aligned}
\E_{\XSR}\left(\|\U - \XSR\|^2\right)
& = \|\U-\La^T\ULR\|^2 +  \Tr(\G - \La^T \A\G\A^T \La) \\
& \geq \|\U-\La^T\ULR\|^2.
\end{aligned}
\end{equation*}
In other words, the kriging component $\La^T\ULR$ has a lower MSE in expectation than the perfect SR samples $\XSR$.
\end{prop}

\noindent The inequality is due to the positiveness of the covariance matrix $\G - \La^T \A\G\A^T \La$ of the innovation component.
This property is consistent with the fact that computing $\E(\X|\XLR)$ aims to minimize the mean squared error.
This is a specific illustration of the well-known regression to the mean problem in SR~\cite{Sonderby2016a}.
The same effect can be observed  for the network SRFlow with temperature $\tau=0$ which solves deterministically the MSE problem~\cite{Lugmayr_et_al_2020}.

\section{Conclusion}

This study solves the problem of stochastic SR for stationary Gaussian textures.
Such texture models constitute a base case for stochastic SR for which all the computations are accessible,
leading to an efficient algorithm for stochastic SR and stochastic SR with a reference image. Our method outperforms some state of the art methods in terms of execution time and LPIPS metrics and we have demonstrated experimentally that the PSNR and the SSIM metrics are not compatible with the evaluation of the perceptual quality of SR samples for textures.
This illustrates the necessity to study further the performance on microtextures of generic stochastic SR models.
Furthermore, the presented method could be applied for other inverse problems involving non invertible operators of the form convolution followed by subsampling.

\bigskip
{\footnotesize \noindent\textbf{Acknowledgements:} The authors acknowledge the support of the project MISTIC (ANR-19-CE40-005).}

\newpage
\bibliographystyle{IEEEbib}
\bibliography{strings.bib,refs.bib}

\end{document}